# Einstein Gravity Path-Integral as an Effective Quantum Theory of a Quantized Space-Time

*Luiz C.L. Botelho*

Departamento de Física
Universidade Federal Rural do Rio de Janeiro
23851–180 – Itaguaí, RJ, Brasil

ABSTRACT

We show that the Einstein Gravitation Theory may be understood as an effective theory of a quantum theory for the space-time implemented as a Bosonic string path-integral and interacting with the fluctuating Einstein space-time metric field (the quantum gravitation vacuum). Additionally, we show how to deduce Yang-Mills Gauge theories from the above cited framework.

In this note we show by means of path-integrals how to deduce Einstein gravitation theory path-integral from the space-time quantized as a Bosonic string wave functional interacting with the Einstein metric field weighted with the cosmological Einstein-Polyakov action.

Let us start our analyze by considering as dynamical fields of our proposed theory, the Einstein gravitational metric $G_{\mu\nu}(X^\alpha)$ and the space-time vector position $X^\alpha(\xi)$, considered as the world-sheet of a closed Bosonic string. The quantum combined system path-integral will supposed to be given by the following $\sigma$-model covariant path-integral which may be regarded as the four-dimensional analogous of similar Path-Integral studied by Polyakov in ref. [1].

$$Z = \int_{(x,\mu,\nu)} \prod d[G_{\mu\nu}(x)] e^{-\mu^2 \int_M (\sqrt{G})(x) d^4 x}$$
$$\left( \int_{\xi \in R^2} \prod (\sqrt{G} G^{\mu\nu}(X(\xi))^{1/2}) dX_\mu(\xi) \right)$$
$$exp\left(-\frac{1}{2\pi\alpha'}\right) \int_{R^2} d^2\xi [\sqrt{G} G_{\mu\nu}(X^\alpha(\xi))(\partial_a X^\mu \partial^a X^\nu)] \quad (1)$$

Let us show that at the string scale $\alpha' \to 0$ (a (small) quantum piece of our classically observed space-time manifold $M$), the string path-integral in eq. (1) leads to an effective (formal) Path-Integral for metric fields $G_{\mu\nu}(x)$ weighted by the Einstein action.

In order to show this we take a (classical) fixed point $\bar{X}^\alpha$ and consider the geodesic expansion around it but defined by the quantum string vector position until the $(X')^2$ order namelly

$$G_{\mu\nu}(X^\alpha(\xi)) = \eta_{\mu\nu}(\bar{X}^\alpha) + \frac{(\alpha')^2}{3} R_{\mu\alpha\nu\beta}(\bar{X}^\alpha) Y^\alpha(\xi) Y^\beta(\xi) + O(\alpha'^4) \quad (2)$$

$$D_{\sqrt{G}} X^\alpha(\xi) = \prod_{\xi \in R^2} dX^\alpha(\xi)[(\sqrt{G} G^{\mu\nu})^{1/2}(X^\beta(\xi))] \sim \prod dY^\alpha(\xi)$$
$$+ O(\alpha'^2) = D^F[Y^\alpha(\xi)] \quad (3)$$

As a consequence of the above made remarks we thus, should consider the following path-integral at the space-time (quantum) chart of Hausdorf dimension four $V_{\bar{x}} = \{X^\alpha + \frac{1}{\sqrt{\alpha'}} Y^\alpha(\xi)\} C R^4$

$$Z_{\bar{V}\{x^\alpha\}}[G_{\mu\nu}(\bar{X}^\alpha)] = \int D^F[Y^\mu(\xi)] \, exp\left\{-\frac{1}{2} \int_{R^2} d^2\xi \sqrt{G}(\bar{X}^\alpha) \right.$$
$$\left. [\eta_{\mu\nu}(\bar{X}^\alpha) + \frac{(\alpha')^2}{3} R_{\mu\alpha\nu\beta}(\bar{X}^\alpha) Y^\alpha(\xi) Y^\beta(\xi)](\partial_a Y^\mu(\xi) \partial^a Y^\nu(\xi)) \right\} \quad (4)$$

Following closely ref.[2], let us introduce composite fields to write eq. (4) as a gaussian path-integral over the string vector position $Y^\beta(\xi)$

$$Z_{V\{\bar{x}\}}[G_{\mu\nu}(\bar{x}^\alpha)] = \int D^F[Y^\mu(\xi)]D^F[\lambda(\xi)]$$
$$exp\left\{-\frac{1}{2}\int d^2\xi\sqrt{G(\bar{X}^\alpha)}\left[(Y^\mu(x))(-\partial_\xi^2)\eta_{\mu\nu}(\bar{X}^\alpha)Y^\nu(\xi))+\right.\right.$$
$$\left.\left.+\frac{1}{3}(\alpha')^2 R_{\mu\alpha\nu\beta}(\bar{X}^\alpha)\sigma^{\alpha\beta}(\xi)(\partial_a Y^\mu \partial^a Y^\nu)(\xi)\right]\right\}$$
$$exp\left\{i\int_{R^2}d^2\xi\sqrt{G(\bar{X}^\alpha)}\lambda(\xi)G_{\alpha\beta}(\bar{X}^\alpha)(\sigma^{\alpha\beta}(\xi)-Y^\alpha(\xi)Y^\beta(\xi))\right\} \quad (5)$$

At this point, we make a departure from the usual wisdom of quantum phases of $\sigma$-models and scale invariance imposed on strings quantum motions on classical backgrounds ([3]),([4]). In this usual context the external field $G_{\mu\nu}(\bar{X}^\alpha)$ is not a dynamic of degree of freedom and is postulated to be a vertex operator of spin-two strings excitations. Note that this vertex interpretation is possible only for weak external $G_{\mu\nu}(\bar{X}^\alpha)$ fields, in order to allow a postulated harmonic oscillator quantization of the $\sigma$-model string and its associated perturbative *ultra-violet* analysis ([4]). We make, thus, the new hypothesis about the existence of a *non-perturbative Kösterlitz-thouless phase* where we have the following non-vanishing world-sheet condensates signaling the breakdown of the $\sigma$-model conformed invariance ([2],[5])

$$\lambda_{\alpha\beta}(\xi) = i<\lambda>G_{\alpha\beta}(\bar{X}) \quad (6)$$

$$\sigma^{\alpha\beta}(\xi) = <\sigma>G^{\alpha\beta}(\bar{X}) \quad (7)$$

Here $<\lambda>$ and $<\sigma>$ are constants condensates with the first value condensate very high, namely $<\lambda>\to\infty$. Substituting eq. (6)-eq.(7) into eq.(5) we get a gaussian functional integral with the following result

$$Z_{V\{\bar{x}\}}[G_{\mu\nu}(\bar{X}^\alpha)] = \lim_{\substack{\alpha'\to 0 \\ <\lambda>\to\infty}}$$
$$\left\{det_F^{-1/2}\left[(\eta_{\mu\nu}(\bar{X})+\frac{<\sigma>}{3}(\alpha')^2 R_{\mu\alpha\nu\beta}(\bar{X})G^{\alpha\beta}(\bar{X})(-\partial^2)_\xi+<\lambda>G_{\mu\nu}(\bar{X})\right]\right\} \quad (8)$$

or equivalently

$$Z_{V\{\bar{x}\}}[G_{\mu\nu}(\bar{x}^\alpha)] = \lim_{<\lambda>\to\infty}\lim_{\alpha'\to 0}\left\{det_F^{-1/2}[(-\partial^2)_\xi\ \eta_{\mu\nu}(\bar{X})+\right.$$
$$\left.+<\lambda>G_{\mu\rho}(\bar{X})(\eta_{\rho\nu}(\bar{X})-\frac{<\sigma>}{3}(\alpha')^2 R_{\rho\alpha\nu\beta}(\bar{X})G^{\alpha\beta}(\bar{X})]\right\} \quad (9)$$

In order to compute the above written functional determinant at the limit of large $<\lambda>$, we regulate this determinant by the proper-time method and evaluate the $<\lambda> \to \infty$ limit as in ref. [6].

We, thus, obtain the Einstein-Hilbert action (in the Euclidean world) in the space-time quantum chart $V\{\bar{x}\}$, namelly

$$Z_{V\{\bar{x}\}}[G_{\mu\nu}(\bar{x}^\alpha)] = \sim \ exp\left\{-\frac{1}{16\pi g}(R(G^{\mu\nu}(\xi)).\sqrt{G(\bar{x})}) + O((\bar{\alpha}')^2)\right\} \tag{10}$$

It is worth point out that the Newton gravitational constant $g$ in our approach is not fundamental in our theory and is defined in terms of the microscopic string condensate constants as follows

$$\frac{1}{16\pi g} = <\sigma>(\alpha')^2.c(\infty) \tag{11}$$

where $c(\infty)$ is an infinite constant regularization scheme dependent and coming from the evaluation of the functional determinant (In the proper-time technique, we have that $c(\infty) = \lim_{\varepsilon \to 0}(\varepsilon^{-P}A)$ with $A$ denoting the internal string area $\varepsilon$ the two-dimensional proper-time parameter with $P > 0$).

The complete path-integral eq. (9) is finally given by sum of eq. (10) over all space-time charts $V_{\{\bar{x}\}}$ with $\bar{x} \in R^4$

$$Z = \int \prod_{x,\mu\nu} d[G_{\mu\nu}(\bar{X})^\alpha] exp\{-\mu^2 \int_M (\sqrt{G})(\bar{X})d^4\bar{X}^\alpha \sum_{\{\bar{X}\}\in R^4} \{Z_{V\{\bar{X}_\alpha\}}\}[G_{\mu\nu}(\bar{X}^\alpha)] \}$$

$$= \int \prod_{(x,\mu,\nu)} d[G_{\mu\nu}(\bar{X}^\alpha)] exp\{-\mu^2 \int_{R^4}(\sqrt{G})(\bar{X}^\alpha)d^4\bar{X}^\alpha\}$$

$$exp\left\{-\frac{1}{16\pi g}\int_{R^4} d^4\bar{X}^\alpha(\sqrt{G}R(G^{\alpha\nu}))(\bar{X}^\alpha)\right\} \tag{12}$$

At this point we remark that the Quantum Field Theory must be quantized at a scale $\ell > 10^{-33}cm$ ([7]) independent of possessing an effective nature.

This is the main result of this paper. For completeness of our study, let us show that our proposed framework may be straightforward generalized to include Gauge Fields associated to compact symmetry groups (see refs. [7]) by imposing the gauge invariant geodesic expansion $A_\mu(\bar{x}^\beta) = \frac{1}{2}y^\rho F_{\rho\mu}(\bar{x}^\beta)y^\alpha$ in the $\sigma$-model lagrangean piece written in terms of Fermionic Variables associated to the charge degrees of freedom (without the

metric field!) ([2])

$$S^{int}_{GROUP}[\psi_a, \psi_a, X^\mu, A^i_\mu] = ie \int_{R^2} d^2\xi [A^i_\mu(X^\beta(\xi))(\bar\psi_a \gamma^A (\lambda_i)_{ab} \psi)(\xi)(\partial_A X^\nu)(\xi)] G_{\mu\nu}(x^\beta(\xi)) \tag{13}$$

By re-writing, thus, the action eq. (13) in terms of the non-abelian strength field by making use of the geodesic gauge fixing

$$\begin{aligned} S^{int}_{Group}[\psi_a, \bar\psi_a, X^\mu, A^i_\mu] \\ = ie \left[ \frac{1}{2} X^\rho(\xi) F^i_{\rho\mu}(\bar X^\alpha)(\bar\psi\gamma^a \lambda_i \psi)(\xi)(\partial_a X^\mu(\xi)) \right] \end{aligned} \tag{14}$$

we obtain that the effective action in the space-time quantum chart $V\{\bar x\}$, is exactly given by the limit of large "string condensate" mass scale parameter $m$

$$\lim_{x'\to 0}\langle det_F^{-\frac{1}{2}}\left\{\frac{1}{2\pi\alpha'}\left(-\partial^2\right)_\xi \eta_{\mu\nu}(\bar X^\alpha) + \frac{1}{2}ie\left[F^k_{\mu\nu}(X^\alpha)(\bar\psi\gamma_a\bar\lambda_k\psi)\partial^a_\xi\right]\right\}\rangle_{\psi,\bar\psi}$$
$$=\sim exp\left\{-\alpha'\langle(\bar\psi\gamma_a\lambda_i\psi)(\xi)(\bar\psi\gamma^a\lambda^\ell\psi)(\xi)\rangle_{\psi,\bar\psi} F^i_{\rho\mu}(\bar X^\alpha) F^\ell(\bar X^\alpha) + O(\alpha')\right\} \tag{15}$$

here $\langle\;\rangle_{\bar\psi,\psi}$ denotes the following fermionic action path-integral average ([2],[7])

$$\langle\theta(\psi,\bar\psi)\rangle = \int D^F[\bar\psi_i(\xi)] D^F[\psi_i(\xi)]\theta(\psi_k,\bar\psi_k) exp\left\{-\frac{1}{2}\int_{R^2} d^2\xi\; \bar\psi_i(\xi)(i\gamma^a\partial_a)\psi_i(\xi)\right\} \tag{16}$$

It again worth remark that the Gauge Group coupling constant is not fundamental in our theory and should be defined in terms of the microscopic two-dimensional intrinsinc fermions current-current on the string world-sheet $R^2$

$$\frac{1}{4g^2} = \alpha'\langle(\bar\psi\gamma_a\lambda^i\psi)(\xi)(\bar\psi\gamma^a\lambda_i\psi)(\xi)\rangle_{\psi,\bar\psi} \tag{17}$$

Note that the combined supersymmetric version of eq. (1) plus eq. (6) leads to the usual supergravity versions of the above obtained Bosonic theories.

As a general conclusion, we have proposed an alternative string theory for grand unification without the usual, conceptual and technical complex difficulties (and mistakes!) always encountered in the celebrated Super-String Everything theories ([1]).

## Acknowledgements

Luiz C.L. Botelho was supported by CNPq – Brazil Science Agency.

# Appendix

In this appendix we present our idea to handle the path-integral eq. (1) in the well-known non-renormalizable Fermi-Thirring from fermion theory in three dimensions (or why the metric field $G_{\mu\nu}(\bar{x})$ must be averaged in our eq. (1), opposite to the usual wisdom in strings theories for Grand Unification).

Let us try to define consistently the Fermi-Thirring Theory in $R^3$

$$Z^{(1)} = \int D^F\psi D^F\bar{\psi}\ e^{-\int_{R^3} d^3x[(\bar{\psi}(i\gamma\partial)\psi - \frac{g^2}{2}(\bar{\psi}\gamma^\mu\psi)^2)](x)}\ e^{-\int_{R^3} d^3x(\bar{\psi}\eta+\eta\psi)(x)} \tag{A.1}$$

since this theory is *trivial* (or non-renormalizable by usual $g$-perturbative analysis ([8])), it may be defined as the very low-energy limit of the well-defined massive quantum electrodynamic

$$Z^{(1)} = \lim_{m\to\infty} \int D\psi D\bar{\psi} DA_\mu\ e^{-\int_{R^3} d^3x((\bar{\psi}(i\gamma\partial)\psi)+g\bar{\psi}\gamma^\mu\psi A_\mu)}\delta^{()}(\partial_\mu A_\mu)$$
$$e^{-\int_{R^3} d^3x\ \frac{1}{2}[A_\mu(-\partial^2+m^2)A_\mu]}\ exp(-\int_{R^3} d^3x(\bar{\psi}\eta+\bar{\eta}\psi)). \tag{A.2}$$

At this point we make the following similarity recipe with our metric-string path integrals analysis

$$(\bar{\psi},\psi)(x) \to G_{\mu\nu}(x^\alpha)$$
$$A_\mu(x) \to Y^\mu(\xi)$$
$$D^F[\psi]D^F[\Psi] \to D^{cou}[G_{\mu\nu}(x^\alpha)]$$
$$D^F[A_\mu(x)] \to D^F[Y^\mu(\xi)]$$
$$m \to <\lambda>$$
$$g \to \alpha'$$